\pdfoutput=1

\documentclass[11pt]{article}

\usepackage[preprint]{nlp4MusA}
\usepackage{tabularx}
\usepackage{booktabs}
\usepackage{times}
\usepackage{latexsym}
\usepackage{amsmath}
\usepackage{float}
\usepackage{natbib}
\usepackage[T1]{fontenc}

\usepackage{xcolor}
\providecommand{\ycz}[1]{\textcolor{black}{{#1}}}

\usepackage[utf8]{inputenc}

\usepackage{microtype}

\usepackage{inconsolata}

\usepackage{graphicx}
\usepackage{musicography}
\usepackage{colortbl}
\usepackage{booktabs}
\usepackage{multirow}
\usepackage{makecell}
\usepackage{pifont}
\usepackage{enumitem}
\usepackage{subfigure}

\title{The Interpretation Gap in Text-to-Music Generation Models}

\author{Yongyi Zang$^*$ \\
  Independent Researcher \\
  \texttt{zyy0116@gmail.com} \\\And
  Yixiao Zhang\thanks{Both authors contributed equally.} \\
  Queen Mary University of London \\
  \texttt{ldzhangyx@outlook.com} \\}

\begin{document}
\maketitle
\begin{abstract}

Large-scale text-to-music generation models have significantly enhanced music creation capabilities, offering unprecedented creative freedom. However, their ability to collaborate effectively with human musicians remains limited. In this paper, we propose a framework to describe the musical interaction process, which includes expression, interpretation, and execution of controls. Following this framework, we argue that the primary gap between existing text-to-music models and musicians lies in the interpretation stage, where models lack the ability to interpret controls from musicians. We also propose two strategies to address this gap and call on the music information retrieval community to tackle the interpretation challenge to improve human-AI musical collaboration.

\end{abstract}

\section{Introduction}\label{sec:intro}


In recent years, the field of human-AI music co-creation has experienced significant advancements~\citep{aisongcontest, cosmic, rau2022vi, bougueng2022calliope}. The advent of large-scale text-to-music generation models has played a crucial role in this progress, enabling generating music with good sonic quality and well-defined musical structures~\citep{musicgen, sd1, sd2, musiclm}.

A primary focus of recent research has been to enhance these models through the incorporation of control signals~\citep{cocomulla, jasco, musiccontrolnet, airgen, diffariff}. This has led to significant success in manipulating dynamics, melody, and chord progressions in generated music contents. While precision in following these control signals can still be improved, these developments represent substantial progress. 

\begin{figure}[tb]
    \centering
    
    \includegraphics[width=\linewidth]{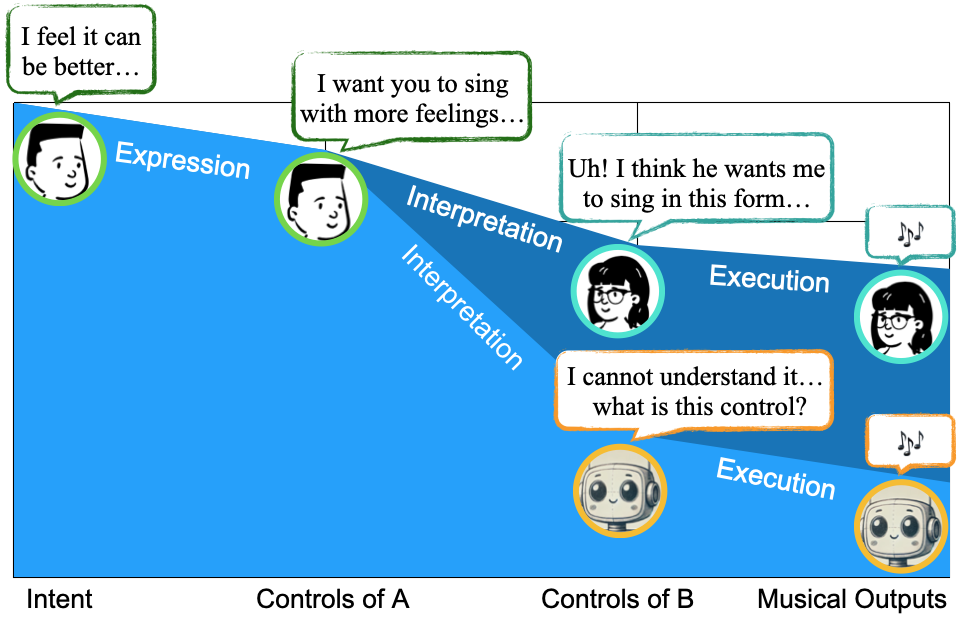}
    \caption{The comparison between human-human and human-AI interaction processes. We observe that the gap exists at both the interpretation stage and the execution stage, while the interpretation stage is often overlooked by current research.}
    \label{fig:example}
\end{figure}


Although \ycz{extensive efforts are made to allow} these models follow control signals \ycz{precisely}, misalignments \ycz{between musicians' intent and model output} still exist, making effective collaboration with musicians challenging~\citep{iteratta,survey1,survey2,tango2}.  In practice, we \ycz{observe} that musicians' control signals require interpretation before execution, and this process is often overlooked in current music information retrieval research. This oversight may hinder the practical applicability of these models in real-world musical settings. Figure~\ref{fig:example} illustrates this issue through a single-round interaction among musician A \ycz{and either musician B or a model}. In this interaction, the control signals expressed by musician A are successfully interpreted by musician B before B generates the musical outputs. In contrast, the model fails to interpret these signals due to the neglected interpretation process in current text-to-music generation models. 


In this paper, our contribution is threefold:

\begin{enumerate}[leftmargin=*, itemsep=0pt]
    \item We propose a framework for the musical interaction process, consisting of three stages: expression, interpretation, and execution of control.
    \item Our literature review identifies a communication gap in current models, which often fail to interpret controls in a way that aligns with human musicians' natural communication methods. 
    \item To address this gap, we propose two approaches: directly learning from human interpretation data or leveraging a strong prior understanding of human interpretation, such as that found in large language models (LLMs).
\end{enumerate}

\begin{table*}[tb]
\scriptsize
    \centering
    \begin{tabularx}{\textwidth}{l|Xp{0.1cm}Xp{0.1cm}Xp{0.1cm}X}
    \toprule
    Cases & Intent & & Controls A & & Controls B & & Outputs\\
    \midrule
    \multicolumn{8}{l}{\textit{Solo Interactions}} \\
    \midrule
    Pianist & Light touch & $\rightarrow$ & Reduce finger force & $\rightarrow$ & N/A & $\rightarrow$ & Piano audio \\
    Experienced Producer & Spacious sound & $\rightarrow$ & Reverb, cut lows & $\rightarrow$ & N/A & $\rightarrow$ & Natural result \\
    Novice Producer & Spacious sound & $\rightarrow$ & Only adding reverb & $\rightarrow$ & N/A & $\rightarrow$ & Unnatural result \\
    Composer & Modulate key & $\rightarrow$ & Write transition & $\rightarrow$ & N/A & $\rightarrow$ & Score \\
    Experienced Guitarist & Emphasizing a chord & $\rightarrow$ & Use complex fingering & $\rightarrow$ & N/A & $\rightarrow$ & Clean strum sound \\
    Novice Guitarist & Emphasizing a chord & $\rightarrow$ & Use complex fingering & $\rightarrow$ & N/A & $\rightarrow$ & Muffled strum sound \\
    \midrule
    \multicolumn{8}{l}{\textit{Multi-Party Interactions}} \\
    \midrule
    Producer \& Experienced Vocalist & Emotive singing & $\rightarrow$ & "More feelings" & $\rightarrow$ & \scriptsize{More dynamics \& articulation} & $\rightarrow$ & Emotional vocal track \\
    Producer \& Novice Vocalist & Emotive singing & $\rightarrow$ & "More feelings" & $\rightarrow$ & \scriptsize{Sing closer to microphone} & $\rightarrow$ & Unnatural vocal track \\
    Experienced Rock Band & Guitar solo & $\rightarrow$ & Gesture & $\rightarrow$ & \scriptsize{Drums and bass play fill; vocalist stop singing} & $\rightarrow$ & Solo section \\
    Novice Rock Band & Guitar solo & $\rightarrow$ & Gesture & $\rightarrow$ & \scriptsize{Everyone ignores the guitarist} & $\rightarrow$ & Solo fights with vocal, creating cacophony \\
    Conductor \& Orchestra & Crescendo & $\rightarrow$ & Rising arms & $\rightarrow$ & \scriptsize{Gradually increasing dynamics} & $\rightarrow$ & Balanced crescendo \\
    DJ \& Crowd & Build energy & $\rightarrow$ & Throwing hands up in the air & $\rightarrow$ & \scriptsize{Crowd thinks it's peak} & $\rightarrow$ & Early climax \\
    \bottomrule
    \end{tabularx}
    \caption{Examples of solo and multi-party musical interactions.}
    \label{tab:diagram_examples}
\end{table*}

\section{Interpretation of Controls}\label{sec:2}

\begin{figure}[htb]
    \centering
\includegraphics[width=\linewidth]{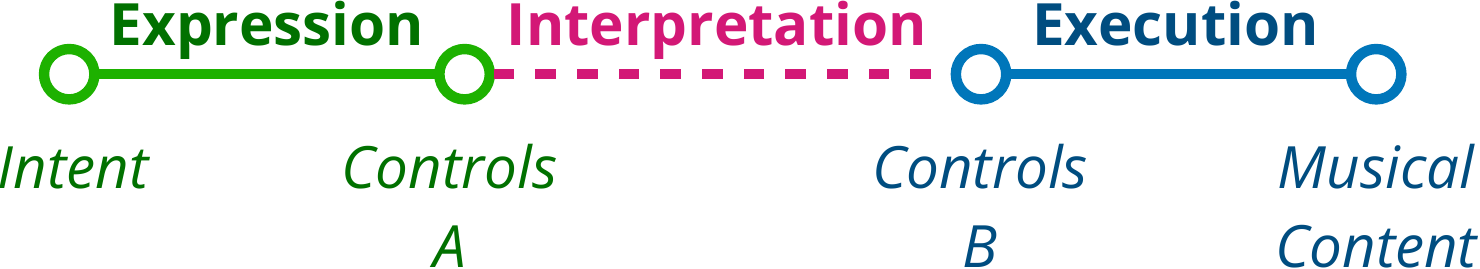}
    \caption{The proposed model that describes musical interaction process. 
    }
    \label{fig:diagram}
\end{figure}

To begin with, we propose a general framework that conceptualizes the musical interaction procedure in three stages: the \textit{expression}, \textit{interpretation}, and \textit{execution} of controls, as shown in Figure~\ref{fig:diagram}. 

In an interaction between parties A and B:

\begin{itemize}[leftmargin=*, itemsep=0pt]
    \item \textbf{Expression}: A's intent is mapped to \textit{Controls A};
    \item \textbf{Interpretation}: B interprets \textit{Controls A}, resulting in \textit{Controls B};
    \item \textbf{Execution}: B executes \textit{Controls B}, producing the final musical output.
\end{itemize}

Table~\ref{tab:diagram_examples} provides several examples illustrating this framework. The framework encompasses both solo and multi-party musical interactions, with the interpretation stage becoming explicit in multi-party scenarios. Successful realization of the original intent hinges on effective mapping across all three stages of the process.

In this section, we examine the musical communication process following this framework. We observe that musical interactions often involve varying degrees of ambiguity in control expression, and skilled musicians can effectively interpret and execute these ambiguous instructions. In contrast, current text-to-music generation models struggle with this ambiguity, and can only understand highly semantical or highly precise instructions.

\subsection{Musicians' Interpretation of Controls}
Musicians communicate through varying levels of ambiguity~\citep{bishop2018collaborative}. The most precise instructions often point to only one outcome (e.g., "Turn the bass 3 dB up") while the most abstract ones require much creative interpretation (e.g., "I want a \textit{moody} synth"). Most communications, however, lie between these two extremes.

Consider this example of a producer addressing a vocalist:~\footnote{Billie Eilish In Studio Making Album "When We All Fall Asleep, Where Do We Go?, \url{https://www.youtube.com/watch?v=Sp-eNvKV0to}} "I want to try one where you just start this \textit{chorus} \textit{very soft}, and in that first phrase, like \textit{[inaudible]}. You know what I mean? \textit{(Sing to demonstrate)} just like, \textit{get crazy} with it. Let's \textit{start quieter} ... or \textit{softer}, or, \textit{babier}. Just try it." This example showcases a wide range of communication types, from highly semantic descriptions (e.g., "very soft," "get crazy," "quieter," "softer," "babier") to performative instructions (e.g., "(Sing to demonstrate)"), and others that fall somewhere in between, requiring interpretation (e.g., "chorus," "first phrase," "[inaudible]"). 


Human musicians excel at interpreting musical instructions with varying ambiguity, a skill known as "musical taste" or "musicianship"~\cite{sloboda1986musical}. This ability enables jazz musicians to adapt improvisations~\cite{berliner2009thinking}, film composers to modify scores for evolving narratives~\cite{cooke2008history}, and orchestral conductors to guide an ensemble through gestures~\citep{bishop2019moving} and verbal cues. This skill, which develops with experience~\cite{lehmann2007psychology}, involves intuitive understanding of musical context, style, and intent~\cite{meyer2008emotion}, allowing musicians to transform ambiguous directions into coherent expressions~\cite{daniel2006your}. 


\subsection{Models' Interpretation of Controls}

\begin{table*}[tb]
\small
    \centering
    \begin{tabular}{lll}
    \toprule
     \textbf{Model}    &  \textbf{Semantic controls} & \textbf{Precise controls} \\
     \midrule
     \multicolumn{3}{l}{\textit{Integrated Controls in Foundation Models}} \\
     Mustango~\citep{melechovsky2024mustango} & Text description, metadata & - \\
     MusicGen~\citep{musicgen} & Text description & melody spectrogram \\
     Diff-A-Riff~\citep{diffariff} & Text description & Music audio mixture \\
     Jen-1 Composer~\citep{jen1composer} & Text description & Other instrument tracks\\
    GMSDI~\citep{gmsdi} & Instrument name & Other instrument tracks\\
     \midrule
      \multicolumn{3}{l}{\textit{Control Enhancement Modules}} \\
     Coco-mulla~\citep{cocomulla} & Text description & Drum track, chord, melody \\
     AIRGen~\citep{airgen} & Text description & Drum track, chord, melody \\
     JASCO~\citep{jasco} & Text description & Drum track, chord, melody \\
      Music ControlNet~\citep{musiccontrolnet} & Text description & Dynamic, melody, rhythm \\
     Jen-1 DreamStyler~\citep{jen1dreamstyler} & Text description & Reference music audio \\
     \midrule
          \multicolumn{3}{l}{\textit{Music Editing Methods}} \\
     MusicMagus~\citep{musicmagus} & Text swapping & Music audio mixture\\
     InstructME~\citep{instructme} & Edit instruction & Music audio mixture\\
      Instruct-MusicGen~\citep{instructmusicgen} & Edit instruction & Music audio mixture\\
    Loop Copilot~\citep{loopcopilot} & Edit instruction & Conversational context (music audio, text)\\
    M$^2$UGen~\citep{m2ugen} & Edit instruction & Conversational context (music audio, text)\\
    ChatMusician~\citep{chatmusician} & Edit instruction & Conversational context (symbolic music, text)\\
    \bottomrule
    \end{tabular}
    \caption{List of representative text-to-music generation models with extra controls. Most controls can be classified into high-level semantic controls and low-level signal-level controls, while the exploration of intermediate-level musicians' communication controls are limited.}
    \label{tab:my_label}
\end{table*}


While human musicians excel at interpreting ambiguous instructions, current music generation models struggle with this task. Traditional approaches to control often rely on disentangling representations in latent space~\citep{luo2019learning, wang2020learning}. For music generative models, control mechanisms are typically implemented through various strategies. Some models integrate controls during initial large-scale pre-training, such as Mustango~\citep{melechovsky2024mustango} and MusicGen~\citep{musicgen}. Others employ post-training model augmentation, exemplified by Coco-mulla~\citep{cocomulla}, AIRGen~\citep{airgen}, and Music ControlNet~\citep{musiccontrolnet}. Additionally, some approaches combine both stages' efforts, as seen in MusicMagus~\citep{musicmagus}, Instruct-MusicGen~\citep{instructmusicgen}, and ChatMusician~\citep{chatmusician}. Despite these advancements in control capabilities, current models still fall short of matching human-level interpretation of nuanced musical instructions.




We posit that the challenge lies not in control implementation methods, but in the nature of the controls themselves. Table~\ref{tab:my_label} summarizes the controls offered by current models, typically either highly semantic (e.g., text descriptions) or highly specific (e.g., chords, melodies). These models struggle with both: for semantic inputs, they mainly interpret at keyword-level rather than understanding natural language~\cite{wu2023audio}, failing with concepts like negation and temporal order~\cite{musiclm, yuan2024t}; for specific inputs, they struggle with precise execution~\cite{instructmusicgen}. When prompts combine semantic and specific instructions, models often fail to interpret the former and fail to execute the latter. The lack of support for other modalities, such as visual cues, also makes effective interpretation more difficult.

While resolving all these challenges is crucial, current research primarily focuses on improving execution ability, such as audio quality, while largely overlooking the interpretation stage. This oversight creates a significant gap in human-AI musical collaboration. Musicians are forced to adapt to the constrained and unnatural controls offered by these models, rather than the models adapting to musicians' natural communication methods. We posit that this mismatch is a key factor in the limited adoption of these otherwise highly capable models by musicians in practice.

\section{Potential Solutions to Improve Interpretation of Controls}\label{sec:3}

Addressing the interpretation gap between musicians and models is challenging due to the complex, multi-modal nature of musician communication, which includes visual cues, textual prompts, vocalizations, and musical references. No existing data sources comprehensively capture all modalities of music interactions, and creating such a dataset would be resource-intensive. Thus, we must approach the problem of learning interpretation under resource constraints. Given these limitations, two potential solutions emerge: directly learning from many aspects of human interpretation data, or leveraging a strong prior understanding of human interpretation, such as that encapsulated in large language models (LLMs). In the following sections, we explore these two avenues for enhancing AI models' ability to interpret musical controls.

\subsection{Directly Learn from Human Interpretation Data}

Previous research has explored many aspects of musical perception and interpretation, including auditory perception~\cite{ananthabhotla2019towards, wright2020perceptual, manocha2020differentiable}, emotion~\cite{yang2012machine, dash2023ai}, song and artist similarity~\cite{knees2013survey, allik2018musiclynx}, music discussions~\cite{hauger2013million}, recommendation systems~\cite{bertin2011million}, and non-verbal communications, such as gesture and dance movements~\cite{gillian2012gesture, fan2011example}. These studies often rely on crowd-sourced evaluations or public data, achieving good interpretations that can serve as control signals, as demonstrated by~\citet{huang2024dance} in music generation from dance movements.

Learning directly from these diverse sources require combining them into cohensive controls, which may be achieved through pseudo-description generation, an approach that has shown promise in music captioning~\citep{mei2024wavcaps, doh2023lp} and understanding~\cite{liu2024music}. 

\subsection{LLMs for Musical Interpretation}



LLMs' robust language understanding enables the decomposition of user queries into specialized tasks, an approach pioneered by HuggingGPT~\citep{hugginggpt}. This method has inspired audio domain projects such as Loop Copilot~\citep{loopcopilot}, WavJourney~\citep{liu2023wavjourney}, WavCraft~\citep{liang2024wavcraft}, and MusicAgent~\cite{yu2023musicagent}. \citet{listen} explores synthesizing natural language from control parameters for model training. 

However, user studies~\citep{analysis, survey1, survey2, loopcopilot} reveal that professional musicians often experience misalignment between model interpretations and their intentions, primarily due to LLMs' lack of domain-specific musical knowledge~\citep{bench1}. Research in other domains indicates that simply integrating domain knowledge can significantly enhance LLMs' capabilities~\cite{lee2024towards}. Consequently, we posit that by collecting domain knowledge and natural music conversations incorporating this knowledge, we could effectively boost LLMs' ability to execute music tasks. Furthermore, these enhanced LLMs could potentially generate synthetic training data for developing more compact interpretation models.

\section{Conclusion}
We identify a critical gap in text-to-music generation models: their inability to effectively interpret musicians' controls. We propose a three-stage framework for musical interaction: expression, interpretation, and execution, and highlight how current AI models often struggle with the crucial interpretation stage. To address this gap, we suggest two potential solutions: directly learning from various sources of human interpretation data and leveraging large language models for musical interpretation. We call on the MIR community to prioritize research in this area, as improving the interpretation capabilities is crucial for their integration into creative workflows and for realizing their full potential as collaborative tools for musicians.

\section*{Ethics Statement}
Our work includes YouTube video transcript excerpts demonstrating artists' creative processes, used solely to illustrate our proposed framework. We thank these amazing artists for sharing their creative processes. All copyrights remain with the original video owners, and excerpts are included for research purposes only.

We acknowledge that musical communications and interpretations encapsulate diverse musicianship, tastes, and cultural nuances. While some aspects of musical communications may be universal, they are often influenced by social culture and individual experiences. We encourage the community to be mindful of this diversity when modeling musical interpretations, as capturing these nuances can enhance the music creation process with generative models.


\bibliography{nlp4MusA}

\end{document}